\documentclass[final,5p,times,twocolumn,authoryear]{elsarticle}

\usepackage{amsmath,amssymb,booktabs,tabularx}
\usepackage{stfloats}      
\usepackage{placeins}
\usepackage[authoryear,round]{natbib}

\usepackage{hyperref}

\usepackage[utf8]{inputenc}
\usepackage{newunicodechar}
\newunicodechar{−}{\ensuremath{-}}

\usepackage{url}
\usepackage{doi}
\usepackage{siunitx}  
\usepackage{graphicx} 
\usepackage{rotating} 
\usepackage{threeparttable} 

\usepackage{xcolor}
\usepackage[normalem]{ulem}
\usepackage{xspace}

\definecolor{editcolor}{RGB}{255,0,0}

\journal{High Energy Astrophysics}

\begin{document}

\begin{frontmatter}

\title{Interstellar Scintillation of Three Nearby Pulsars with FAST}

\ead{dangsj@gznu.edu.cn; wuzw@bao.ac.cn}

\author[aff1]{Ying-ying Ren}
\author[aff1]{Shi-jun Dang\corref{cor1}\fnref{fn1}}
\author[aff2]{Zi-wei Wu\corref{cor1}\fnref{fn2}}
\author[aff3,aff4]{Yu-lan Liu}
\author[aff2,aff5]{Yan-qing Cai}
\author[aff1,aff6]{Qi-jun Zhi}
\author[aff1]{Lun-hua Shang}
\author[aff1]{Ru-shuang Zhao}

\cortext[cor1]{Corresponding author}
\fntext[fn1]{ORCID: \href{https://orcid.org/0000-0002-2060-5539}{0000-0002-2060-5539}}
\fntext[fn2]{ORCID: \href{https://orcid.org/0000-0002-1381-7859}{0000-0002-1381-7859}}

\affiliation[aff1]{
  organization={School of Physics and Electronic Science, Guizhou Normal University},
  addressline={Guiyang},
  city={Guizhou},
  postcode={550025},
  country={China}
}

\affiliation[aff2]{
  organization={State Key Laboratory of Radio Astronomy and Technology, National Astronomical Observatories, Chinese Academy of Sciences},
  addressline={Beijing},
  postcode={100012},
  country={China}
}

\affiliation[aff3]{
  organization={CAS Key Laboratory of FAST, National Astronomical Observatories, Chinese Academy of Sciences},
  addressline={Beijing},
  postcode={100101},
  country={China}
}

\affiliation[aff4]{
  organization={Guizhou Radio Astronomical Observatory, Guizhou University},
  addressline={Guiyang},
  city={Guizhou},
  postcode={550001},
  country={China}
}

\affiliation[aff5]{
  organization={University of Chinese Academy of Sciences},
  addressline={Beijing},
  postcode={100049},
  country={China}
}

\affiliation[aff6]{
  organization={Guizhou University},
  addressline={Guiyang},
  city={Guizhou},
  postcode={550001},
  country={China}
}

\begin{abstract}
Interstellar scintillation probes the properties of the ionized interstellar medium as well as the dynamical behavior of pulsars themselves. Using the Five-hundred-meter Aperture Spherical Radio Telescope, we obtained hour-long observations of PSRs~J0837+0610, J1136+1551, and J1239+2453. We detected a single scintillation arc in PSRs~J0837+0610 and J1239+2453, and identified three distinct arcs in PSR~J1136+1551. Our analysis reveals that the arc curvature scales with observing frequency as $\eta \propto \nu^{-2.0\pm0.6}$ for PSR J0837+0610, and as $\eta \propto \nu^{-1.9\pm0.6}$ for PSR J1239+2453. For PSR J1136+1551, the two clearest arcs exhibit scaling relations of $\eta \propto \nu^{-1.6\pm0.6}$ and $\eta \propto \nu^{-2.0\pm0.4}$, respectively. However, the frequency dependence of the third arc could not be constrained due to its low signal-to-noise ratio at higher frequencies. Moreover, the corresponding scattering screens are measured at distances ranging from 30 to 420 pc from Earth. However, long-term scintillation monitoring or VLBI observations are needed to reliably measure the scattering screen.
\end{abstract}


\end{frontmatter}

\section{Introduction}
\label{sect:intro}

When radio waves from distant compact sources pass through the ionized interstellar medium, fluctuations in its refractive index introduce phase variations. 
The interference of these scattered waves results in a modulation of signal intensity with frequency and time, a phenomenon known as interstellar scintillation \citep[ISS,][]{sch68}. 
Given their small angular sizes, pulsars are typically observed as scintillating radio sources \citep{hbp+68}.
Interstellar scintillation (ISS) is categorized into two primary types: diffractive ISS (DISS) \citep{ric69} and refractive ISS (RISS) \citep{rcb84}. DISS arises from interference effects, producing pronounced intensity modulations on short timescales. In contrast, RISS results from the large-scale focusing and defocusing of radio waves, leading to more gradual modulations over longer timescales \citep{ric90}.

A dynamic spectrum, a two‑dimensional representation of radio signal intensity as a function of time and frequency, enables the measurement of the fundamental scintillation parameters—the scintillation bandwidth ($\Delta \nu_{\rm{d}}$) and scintillation timescale ($\tau_{\rm{d}}$). These parameters serve as primary tools for probing the properties of the ionized interstellar medium and of pulsar systems themselves \citep{ric69}.
At the beginning of this century, the detection of scintillation arcs from the secondary spectrum, that is, the power spectrum of the dynamic spectrum, provided a powerful new tool \citep{smc+01}.
These tools have many applications, for example, mapping the distribution of the Galactic Electron Density \citep{cl02,rmo+25}, constraining the inclination angle of binary systems \citep{lyn84},  resolving pulsar magnetospheres \citep{cwb83, pmdb14}, the proper motion of pulsars \citep{ls82,cor86}, and improving the pulsar timing precision \citep{wksv08}.

The scintillation arc method enables more precise measurements of the scintillation velocity and the location of the scattering screen.
The origin of scattering screens is commonly associated with structures such as the Local Bubble, supernova shells, bow shocks, and H\textsc{ii} regions \citep{yzm+21, mma+22, lmv+23, occ+24}.
These IISM structures can be stratified and studied by analyzing scintillation arcs \citep{rmo+25}.
However, the curvature of a scintillation arc is strongly frequency-dependent \citep{wmsz04,crsc06}. 
As a result, the arcs we observe are frequently broad. 
When the curvatures of two scintillation arcs are similar, it sometimes becomes difficult to distinguish them. Recent applications of the NuT transform \citep{swm+21} and the Normalized Secondary Spectrum \citep{fcm+14,rcb+20} have provided a means to sharpen scintillation arcs, revealing features that were previously hidden. These methods sharpen scintillation arcs by correcting the frequency dependence of the arc curvature—either by resampling the dynamic spectrum in wavelength or by scaling the time axis with frequency—so that the curvature becomes independent of frequency, concentrating the dispersed power into a sharp arc. 
Moreover, to sample the flux variations caused by interstellar scintillation, powerful instruments are required—a role for which the Five-hundred-meter Aperture Spherical Telescope (FAST) is well‑suited to fill \citep{jyg+19,jth+20}.

In this work, we aim to study the ISS properties of bright and nearby pulsars with FAST.
In section~\ref{sect:Obs}, we describe the FAST observations and data processing.
Section~\ref{sect:results} is dedicated to the results.
The conclusion and discussion are encapsulated in Section~\ref{sect:cd}.

\begin{table*}[ht]
  \centering
  \caption{Summaries of FAST observation and scintillation parameters}
  \label{tab1}
  \begin{tabular}{ccccccccc}
    \hline
    \hline
    Name & MJD & $\Delta {\rm{t}}$ & $\Delta {\rm{f}}$ & $T_{\rm{obs}}$ & $\Delta \nu_{\rm{d}}$ & $\tau_{\rm{d}}$ & $\delta_{\rm{est}}$ & $\eta$ \\
         &     & (seconds) & (MHz) & (mins) & (MHz) & (mins)   &  & ($s^{3}$) \\
    \hline 
    J0837+0610 & 59680.5 & 12.7 & 0.122 & 60 & 33.2$\pm$6.8 & 12.3$\pm$2.5 & 0.21   & 0.0384$\pm$0.0038 \\
    J1136+1551 & 58808.1 & 11.2 & 0.122 & 60 & 124.6$\pm$37.8 & 7.1$\pm$2.2 & 0.30  & 0.0020$\pm$0.0003 \\
    -- & -- &  -- &  -- & -- & -- & -- & -- & 0.0040$\pm$0.0004 \\
    -- & -- &  -- &  -- & -- & -- & -- & -- & 0.0127$\pm$0.0014 \\
    J1239+2453 & 58865.9 & 55.0 & 0.122 & 120 & 120.9$\pm$44.4 & 18.7$\pm$6.4 & 0.34 & 0.0296$\pm$0.0018 \\
    \hline 
    \hline
  \end{tabular}
\vspace{0.2cm}

\smallskip
{\footnotesize Note: given are the pulsar name, the modified Julian day, the time resolution, frequency resolution of observations, the derived scintillation bandwidth, scintillation timescale, the statistical error due to the limited number of scintles, and the arc curvature.}
\end{table*}

\begin{figure*}[ht]
    \centering
    \includegraphics[width=0.33\linewidth]{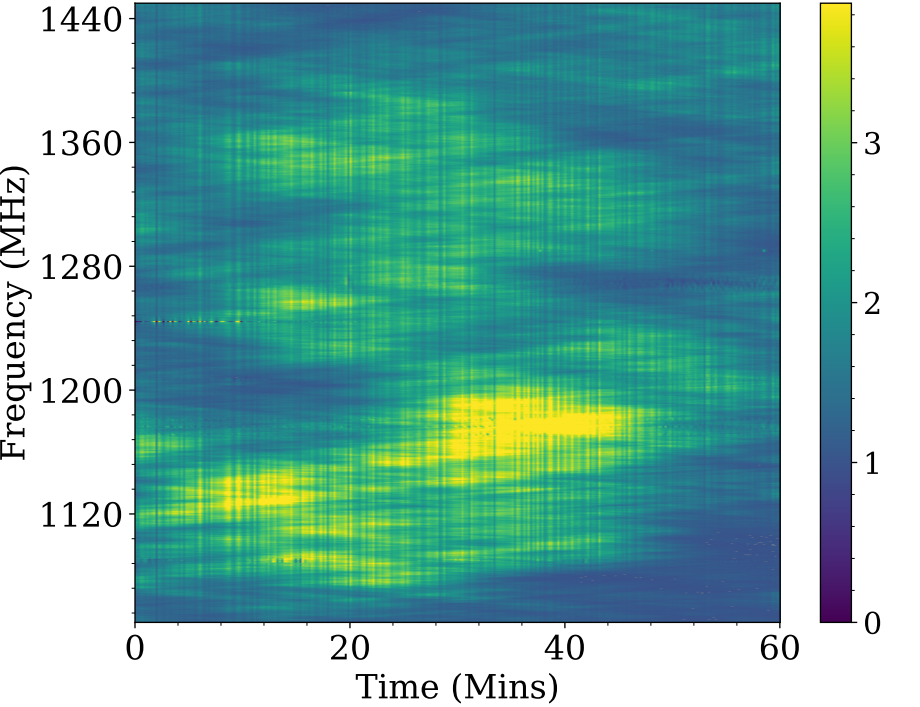}
    \includegraphics[width=0.33\linewidth]{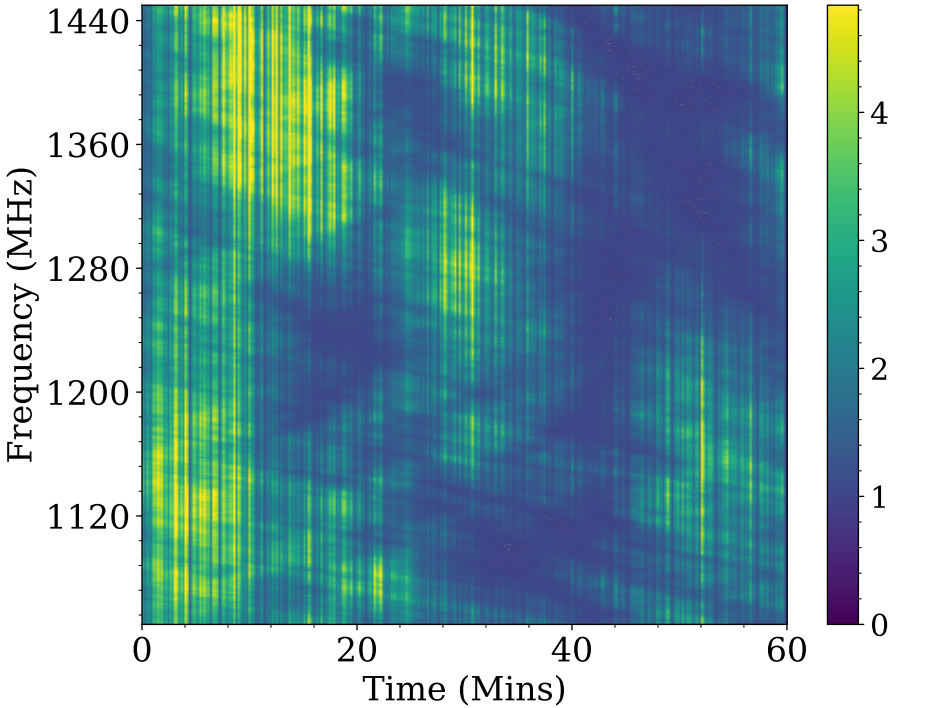} 
    \includegraphics[width=0.33\linewidth]{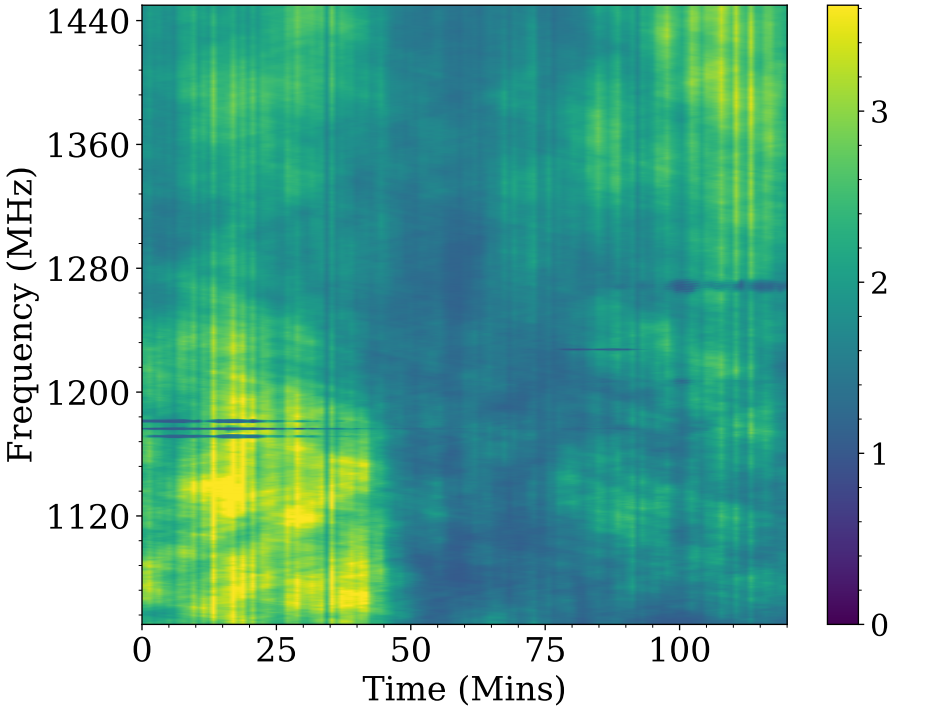}
    \caption{The dynamic spectra of PSR~J0837+0610 (left), J1136+1551 (middle), and J1239+2453 (right) with FAST. These vertical streaks visible in the secondary spectrum of PSR~J1136+1551 are due to the nulling behavior (see Figure~\ref{fig:nulling}).}
    \label{fig:ds}
\end{figure*}

\section{Observation and Data Processing}
\label{sect:Obs}

Based on pulsar flux density and dispersion measure, we selected PSR~J1136+1551, PSR~J0837+0610, and PSR~J1239+2453 from the released FAST data set \footnote{https://fast.bao.ac.cn}. 
The observations were conducted using FAST's 19-beam receiver in the L-band (\SIrange{1.0}{1.5}{\GHz}).
The data were recorded with a time resolution of \SI{49.152}{\micro\second} and 4096 frequency channels.
The data processing pipeline included: initial folding and dedispersion of raw data using DSPSR (\citealt{vb11}); sub-integration times set to \SI{10}{\second} based on the nulling ratio of PSR~J1136+1551 to optimize scintillation arc analysis. 
Radio Frequency Interference (RFI) from terrestrial transmitters, radar, and satellites was automatically identified and excised using the PAZ (Pulsar Archive Zapper) module in PSRCHIVE (\citealt{hvm04}). 
Remaining RFI was manually removed using the Pulsar Archive Zapper Interface (PAZI) module. 
The initial dynamic spectra were produced using the PSRFLUX tool in PSRCHIVE \citep{hvm04}.

We corrected for the instrumental bandpass, as well as the pulse-to-pulse and flux variations at different frequencies of the pulsar that are attributable to the emission mechanism. The two-dimensional autocorrelation function (2D ACF) of the obtained dynamic spectrum is then computed. Following the method of \citet{cmr+05} and \citet{rcn+14}, we fit one-dimensional cuts of the 2D ACF  to derive the scintillation bandwidth ($\Delta \nu_{\rm{d}}$) and scintillation timescale ($\tau_{\rm{d}}$), as displayed in Figure~\ref{fig:acf} of the Appendix~\ref{Appendix}.

To reduce aliasing effects in the secondary spectrum, we first applied a Hamming window to the outer 10\% of each dynamic spectrum. 
The NuT transform was then applied to these windowed spectra \citep{vv22} to obtain the secondary spectra, i.e., the power spectra of the dynamic spectra.
The data processing pipeline and scintillation parameter extraction closely follow the methodology described by \cite{wvm+22}. 
The scintillation arc curvature was measured by summing the power at all points corresponding to a given curvature $\eta$ in the secondary spectrum \citep{bot+16}, {as displayed in Figure~\ref{fig:c} of the Appendix~\ref{Appendix}.
This yields an intensity–curvature relation, the peak of which was determined by fitting a Gaussian function, with the uncertainties derived from the standard deviation (sigma) of the Gaussian fit.

\section{Results}
\label{sect:results}

The dynamic spectra of three selected pulsars are shown in Figure~\ref{fig:ds}. Flux variations caused by interstellar scintillation (ISS) and criss-cross patterns are clearly visible, indicating the presence of scintillation arcs.
The scintles are well resolved, as both the frequency and time resolutions of the dynamic spectra are significantly finer than the characteristic scales of scintles in the frequency and time domains.
We extracted the basic scintillation parameters—the scintillation bandwidth $\Delta \nu_{\rm{d}}$ and the scintillation timescale $\tau_{\rm{d}}$—which are listed in Table~\ref{tab1}. Their relatively large uncertainties are due primarily to the limited number of scintles available in the observations. 
The measured scintillation bandwidths $\Delta \nu_{\rm{d}}$ of the three pulsars are consistent with earlier measurements \citep{wvm+22}, under the assumption of a Kolmogorov turbulence spectrum and considering RISS modulation \citep{brg99, lvm+22}.

The secondary spectra for the three pulsars are shown in Fig.~\ref{fig:ss}. We note that the observed vertical striations in the secondary spectrum of PSR~J1136+1551 are attributed to the pulsar’s nulling behavior, as shown in Figure~\ref{fig:nulling} in Appendix~\ref{Appendix}. We attempted to decrease the time resolution to suppress the effects of nulling; however, this smeared the scintillation features.
Singe-pulse-based ISS method will be tested \citep{wzf+25} on nulling pulsars.  
Each of PSR~J0837+0610 and PSR~J1239+2453 exhibits a single scintillation arc, whereas three distinct arcs are clearly seen in the spectrum of PSR~J1136+1551. 
While previous studies of PSR~J0837+0610 have shown that its secondary spectrum exhibits a single arc with two subcomponents on it \citep{bmg+10}, our data suggest the possible presence of a second arc, though its detection is only marginal. 
We note that the detection of two closely located scattering screens for this pulsar has also been reported by \cite{bbv+23} using VLBI data.\cite{wvm+22} did not detect a scintillation arc for PSR~J1239+2453 at around 150~MHz, which may be due to the small differential time delay $f_{\rm{t}}$ and large scintillation arc curvature at their frequency ($\eta_{\rm{150~MHz}}$ $\sim$ 2.1~$s^{3}$ assuming $\eta \propto f^{2}$).

\begin{figure*}[ht]
    \centering
    \includegraphics[width=0.33\linewidth]{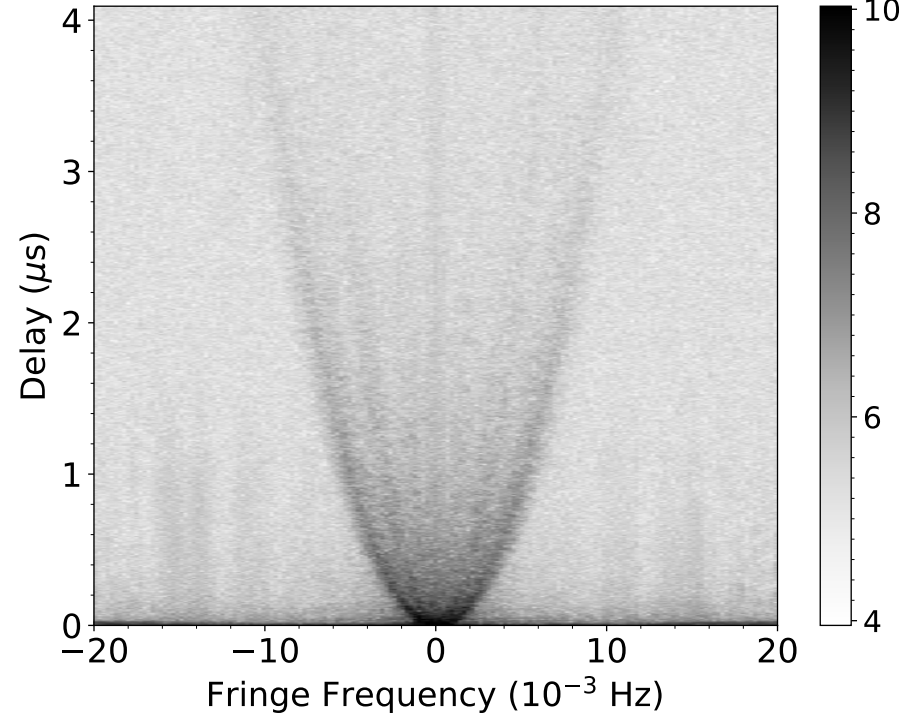}
    \includegraphics[width=0.33\linewidth]{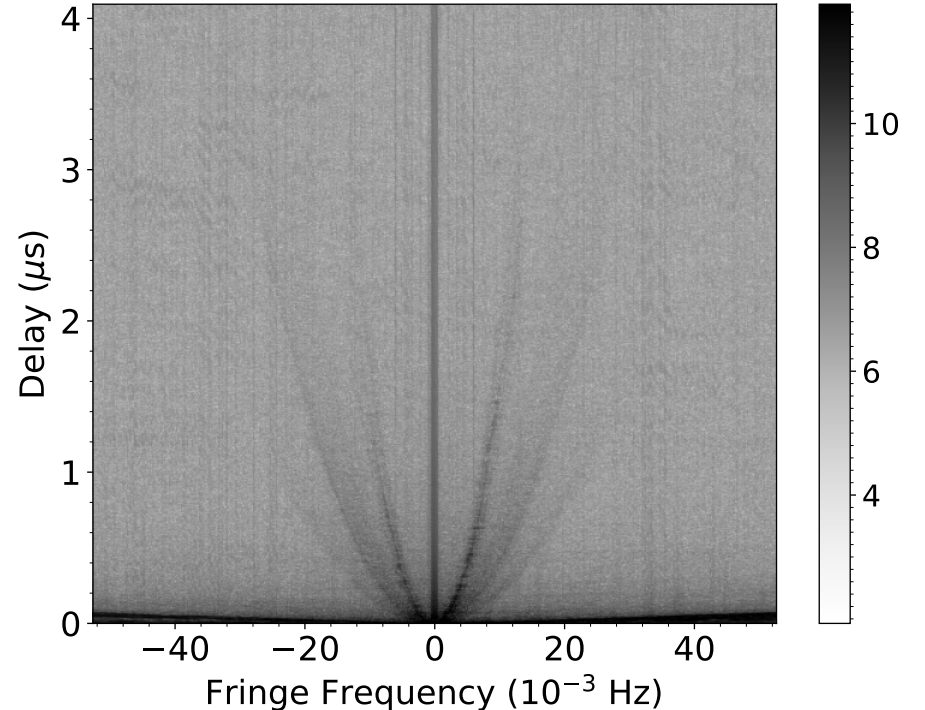} 
    \includegraphics[width=0.33\linewidth]{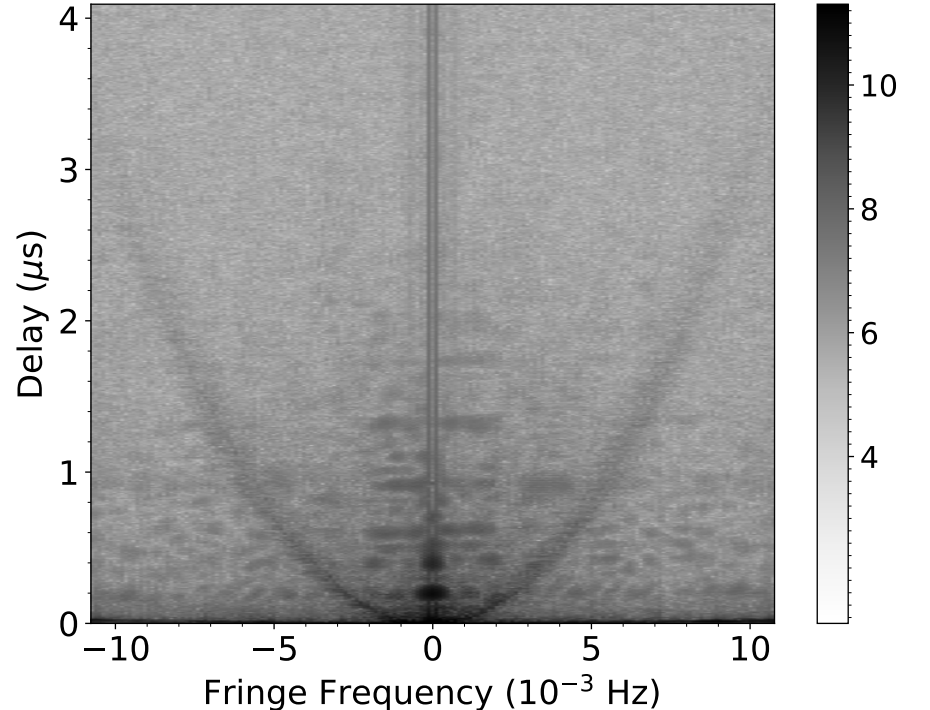}
    \caption{The secondary spectra of PSR~J0837+0610 (left), J1136+1551 (middle), and J1239+2453 (right) with FAST. We applied the NuT transform to sharpen the scintillation arcs.}
    \label{fig:ss}
\end{figure*}

The corresponding arc curvatures are summarized in Table~\ref{tab1}.
The parabolic scintillation arc in the secondary spectrum is defined by the relation \citep{smc+01}:
\begin{equation}
f_{\rm{d}} = \eta f_t^2,
\label{eq:parabolic_arc}
\end{equation}
where $f_{\rm{d}}$ is the differential Doppler shift, $f_t$ is the differential time delay, $\eta$ is the curvature parameter of the scintillation arc. 
The arc curvature is given by \citep{wmsz04,crsc06}:
\begin{equation}
\eta = \frac{D s (1 - s)c}{2 f^2V_{\mathrm{eff}}^2 \cos^2 \psi},
\label{eq:curvature_parameter}
\end{equation}
where $D$ is the pulsar distance, the fractional screen $s$ ranges from 0 (at the pulsar) to 1 (at the observer), c is the speed of light, $f$ is the observingfrequency and $\psi$ denotes the angle between the scintillation velocity $V_{\mathrm{eff}}$ and the major axis of the anisotropic scattering screen. 
The effective scintillation velocity is the combination of pulsar $V_{\rm{p}}$, earth $V_{\rm{E}}$ and screen $V_{\rm{ISM}}$ velocities \citep{rch+19}. 
\begin{equation}
V_{\mathrm{eff}} = (1 - s)V_{\rm{p}} + sV_{\rm{E}} - V_{\rm{ISM}}.
\label{eq:velocity}
\end{equation}

We assume an isotropic and stationary scattering screen ($\psi$ = 0, and $V_{\rm{ISM}}$ = 0).
The Earth's velocity $V_{\rm{E}}$ is then calculated \footnote{https://github.com/danielreardon/scintools} \citep{rcb+20}.
The pulsar distances are 620$\pm$60~pc for J0837+0610 \citep{lpm+16}, 372$\pm$3~pc for J1136+1551 \citep{dgb+19}, and 850$\pm$60~pc for J1239+2453 \citep{bbg+02}. We estimate the scattering screen for PSR~J0837+0610 to be located at a distance of  421$\pm$60~pc from Earth. To facilitate a direct comparison, we recalculated the scattering screen distance for PSR J0837+0610 using the same scattering screen velocity, anisotropy angle, and pulsar distance adopted by \citet{bmg+10}. Our derived distance is 413$\pm$66 pc from Earth.

This is consistent with  \cite{bmg+10}, suggesting that the same scattering screen has dominated the observed interstellar scintillation over decades. 
The three scattering screens of PSR~J1136+1551 are located at distances of \(196\pm14\)~pc, \(146\pm11\)~pc, and \(39\pm5\)~pc from Earth, respectively. 
By visual inspection of the secondary spectra, this is consistent with the findings of \cite{mpj+23} and corresponds to screens B, C, and E among the six scintillation arcs reported by \cite{mzs+22}.
For PSR~J1239+2453, the scattering screen is located at a distance of \(253\pm36\)~pc from Earth.

\begin{figure*}[ht]
    \centering
    \includegraphics[width=0.38\linewidth]{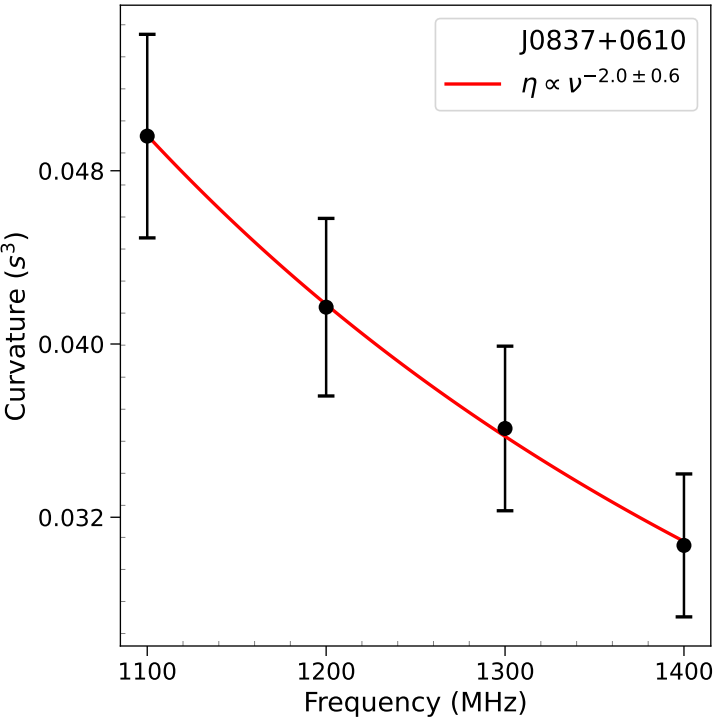}
    \includegraphics[width=0.38\linewidth]{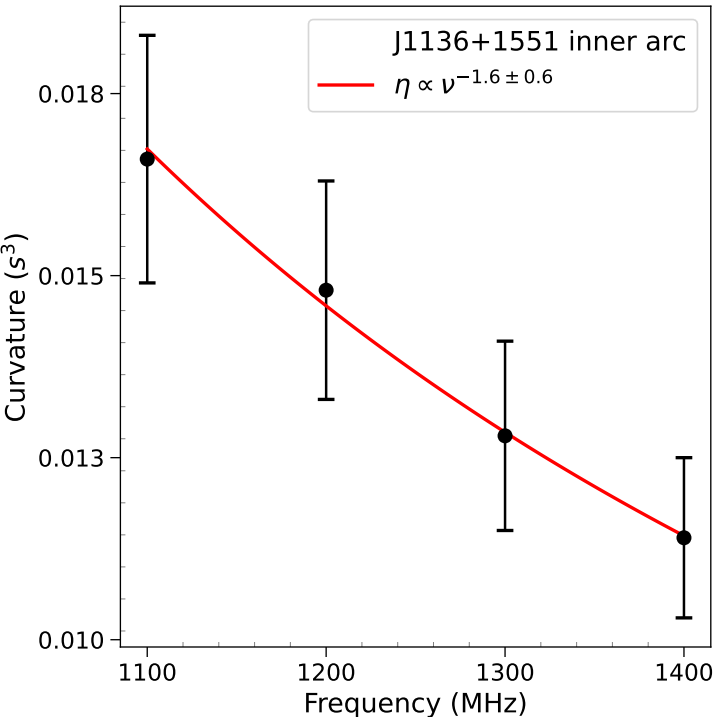}  \\
        \includegraphics[width=0.38\linewidth]{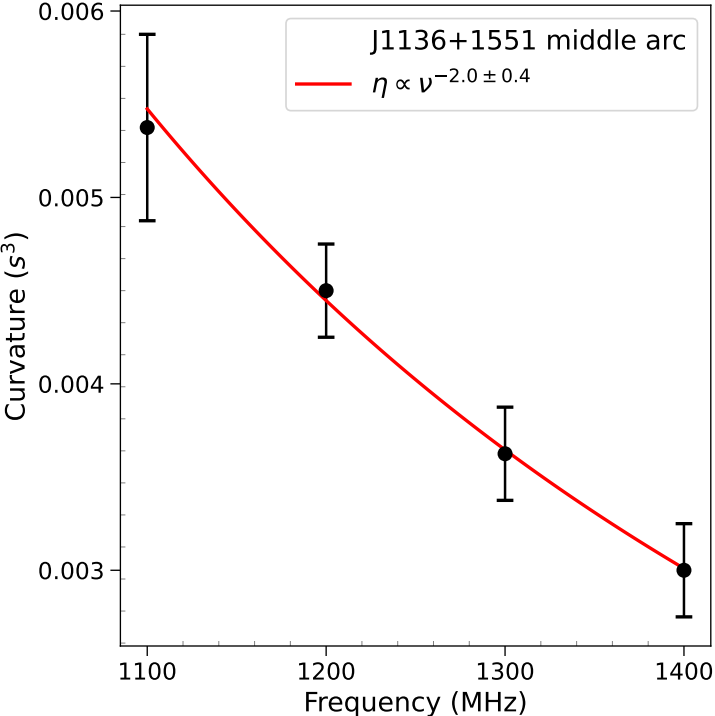}
    \includegraphics[width=0.38\linewidth]{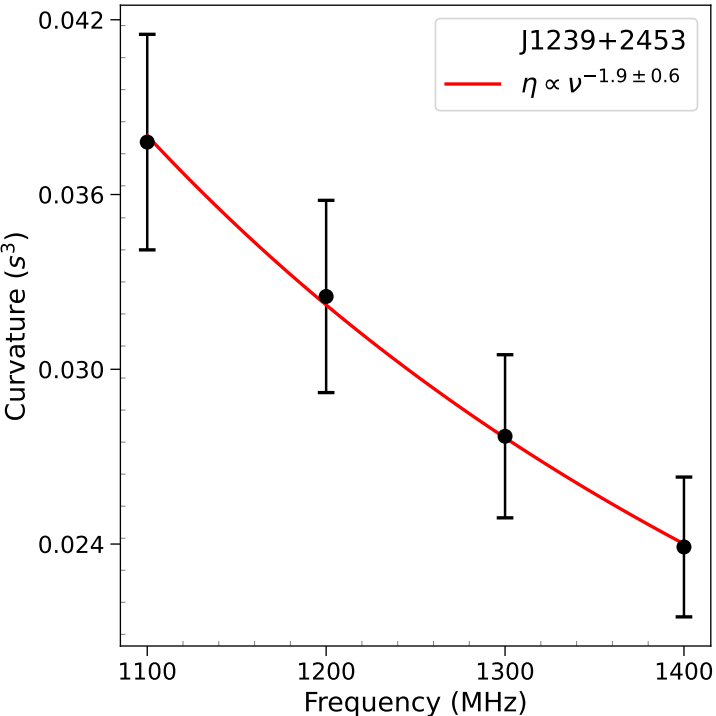} 
    \caption{The scintillation arc curvature as a function of frequency of PSRs~J0837+0610, J1136+1551, and J1239+2453 with FAST.}
    \label{fig:af}
\end{figure*}

The curvature of scintillation arcs exhibits a strong frequency dependence \citep{hsb+03}. 
We divided our 400~MHz bandwidth into four contiguous 100~MHz subbands, centered at 1100, 1200, 1300, and 1400~MHz, respectively. 
Figure~\ref{fig:af} presents the measured scintillation-arc curvatures as a function of frequency. The outer arc of PSR~J1136+1551 could not be detected in all sub‑bands; therefore, its frequency dependence could not be studied.
The derived power-law indices are consistent with the predictions of a simple thin scattering screen model \citep{wmsz04,crsc06}. The larger uncertainties in these measurements primarily stem from the limited fractional bandwidth of FAST (400/1250 = 0.32).

\section{Conclusion and Discussion}
\label{sect:cd}
We have investigated the ISS properties of three nearby, bright pulsars with FAST. For PSR~J0837+0610 and PSR~J1239+2453, a single dominant scattering screen was detected; for PSR~J1136+1551, three distinct screens were identified, confirming the multi‑screen structure reported in earlier work. 
Since each pulsar was observed in only a single epoch with FAST in this study, the properties of the scattering screens are not fully resolved. Nevertheless, the derived screen distances are broadly consistent with previous measurements. 
With the exception of J0837+0610, the remaining four arcs detected toward the other two pulsars lie within approximately 100–200~pc of Earth, it is possible that they share a common origin related to the Local Bubble. \citep{zga+22}. 
The three arcs detected toward J1136+1551 could similarly be related to substructures within the Local Bubble, analogous to the case of J0437-4715 \citep{rmo+25}. 
A systematic study of nearby pulsars (within 500 pc) therefore appears promising for mapping the detailed structure of the Local Bubble.

Compared to the scintillation bandwidth, scintillation arc curvature offers a more direct constraint on the Kolmogorov spectrum of interstellar turbulence, as it is not affected by statistical uncertainties. 
In this work, the observed frequency scaling of the scintillation-arc curvature is consistent with a Kolmogorov turbulence spectrum. 
While previous studies have reported anomalous frequency scaling in some pulsars, such deviations do not necessarily imply a departure from Kolmogorov turbulence in the IISM. 
Alternative explanations—such as the geometry of the scattering screen—must be considered \citep{cl01,wcv+23}.
This point is particularly relevant because the time-variable pulse arrival delay induced by the IISM is now recognized as one of the major contributors to pulsar timing noise \citep{vs18}. 
By studying ISS properties at lower frequencies, it may be possible to correct for such delays at higher observing frequencies. 
In this approach, the power-law indices of scattering parameters—such as scintillation bandwidth, arc curvature, and scattering timescale—constitute key quantities.

\section*{Acknowledgements}
This work was supported by the Strategic Priority Research Program of the Chinese Academy of Sciences, Grant No.XDA0350501, the College Students' Innovative Entrepreneurial Training Plan Program (Project Number: 2024106631119), the Major Science and Technology Program of Xinjiang Uygur Autonomous Region (No.2022A03013-4 and No.2022A03013-2), the Foundation of Guizhou Provincial Education Department (grant Nos.KY(2023)059), the Guizhou Provincial Basic Research Program (Natural Science) (No. Qiankehejichu-MS(2025)263), and NSFC grant No.12503056. This work made use of the data from FAST (Five-hundred-meter Aperture Spherical radio Telescope)(https://cstr.cn/31116.02.FAST). FAST is a Chinese national mega-science facility, operated by National Astronomical Observatories, Chinese Academy of Sciences.

\bibliographystyle{elsarticle-harv}
\bibliography{journals,modrefs,psrrefs,crossrefs}

\onecolumn

\section{Appendix}
\label{Appendix}

\begin{figure*}[htbp]
    \centering
    \includegraphics[width=0.68\columnwidth]{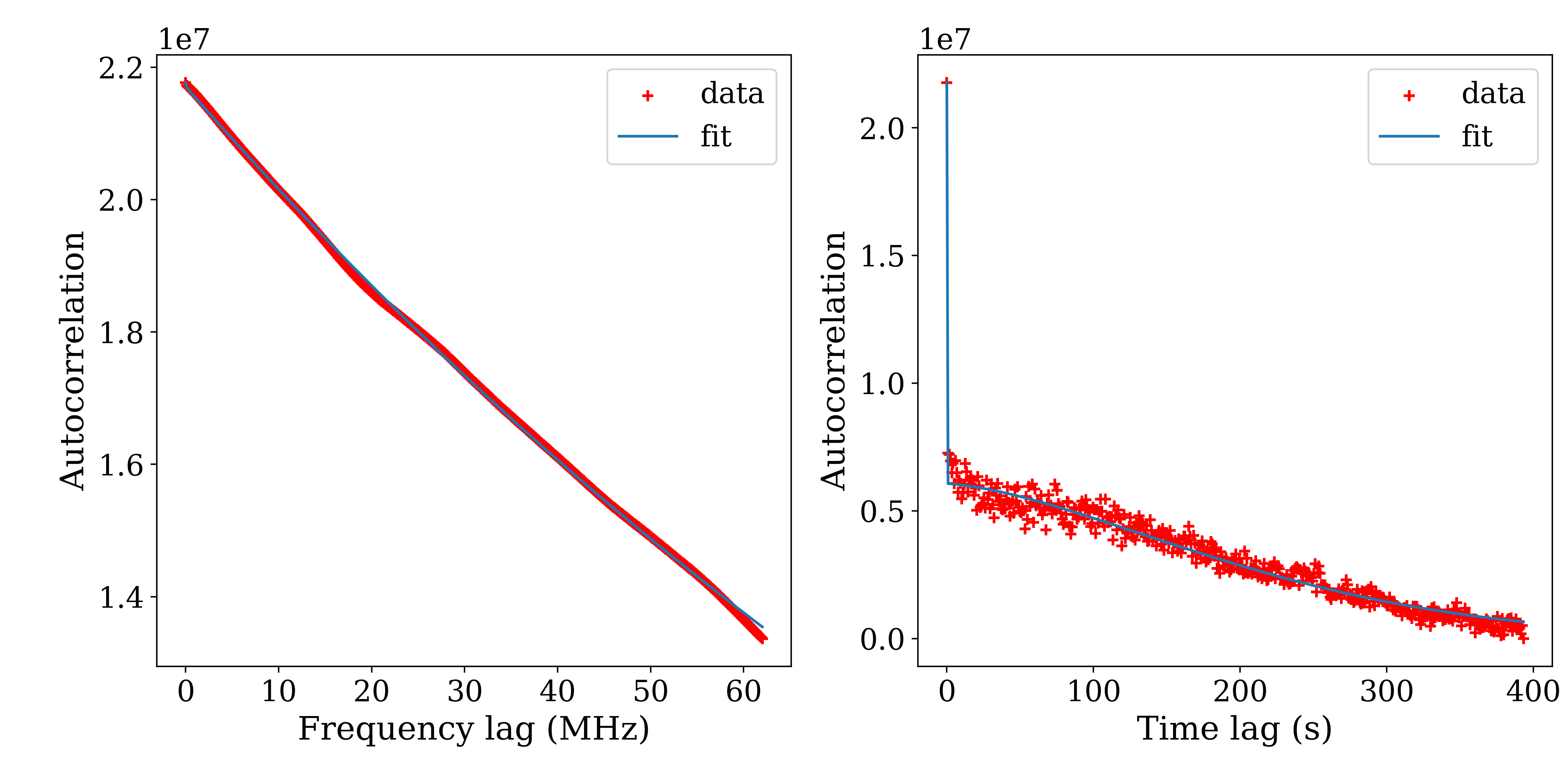}
    \caption{The one-dimensional ACF fits of the dynamic spectrum of PSR~J0837+0610 with FAST in frequency lag (left) and time lag (right).}
    \label{fig:acf}
\end{figure*}

\begin{figure*}[htbp]
    \centering
    \begin{minipage}{0.48\textwidth}
        \centering
        \includegraphics[width=\textwidth]{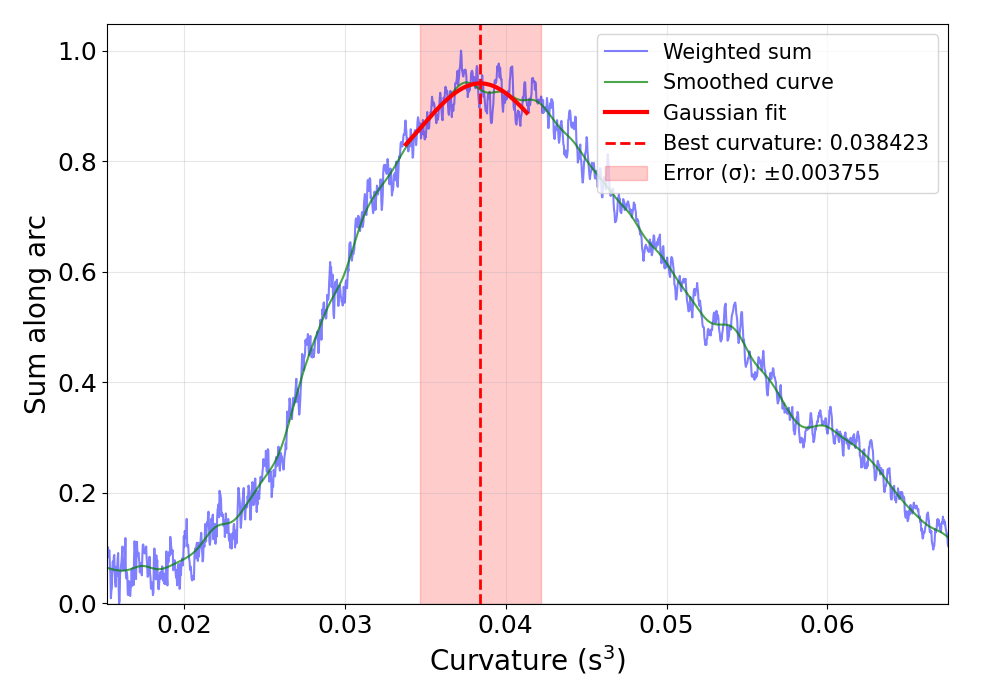}
        \caption{Arc strength as a function of curvature for PSR~J0837+0610. The intensity is summed over all points in the secondary spectrum corresponding to a given curvature $\eta$, following the method of \citet{bot+16}.}
        \label{fig:c}
    \end{minipage}
    \hspace{0.05\textwidth} 
    \begin{minipage}{0.35\textwidth}
        \centering
        \includegraphics[width=0.9\columnwidth]{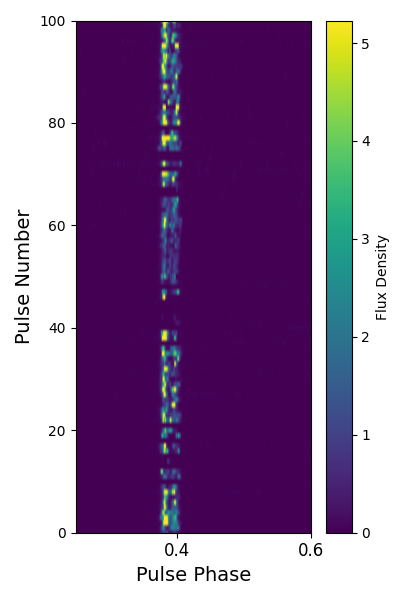}
        \caption{Nulling behaviour of PSR~J1136+1551 with FAST at 1250~MHz. The phase-time plot displays a sample of 100 pulses (selected from the total 3032 pulses) as a function of pulse phase and pulse number. 
        }
        \label{fig:nulling}
    \end{minipage}
\end{figure*}

\end{document}